\documentclass[aps,prd,twocolumn,showpacs,superscriptaddress,nofootinbib,preprintnumbers]{revtex4-2}

\usepackage{amssymb,amsmath,verbatim,mathtools,needspace,enumitem,etoolbox,graphicx,physics,microtype,afterpage,xspace,tabularx,lmodern,multirow}
\usepackage{booktabs}
\usepackage{siunitx}
\usepackage{gensymb}
\usepackage{ulem} 
\usepackage{amsmath}
\usepackage{tikz}
\usepackage{tikz-feynman}
\usepackage{booktabs}
\usepackage{bm}
\usepackage{cancel}
\tikzfeynmanset{
  compat=1.1.0
}
\tikzset{every path/.append style={thick}}
\usepackage{amsfonts}
\usepackage{amssymb}
\usepackage{color}
\usepackage{graphicx}
\usepackage{fontawesome}
\usepackage{xstring}
\usepackage{bm}
\usepackage[colorlinks=true,citecolor=blue,urlcolor=blue]{hyperref}
\hypersetup{colorlinks=true,citecolor=romared,linkcolor=romared,urlcolor=romared}
\definecolor{romared}{RGB}{142,0,28}

\usepackage{afterpage}
\usepackage{multirow}
\usepackage{appendix}
\usepackage[table]{xcolor}
\usepackage{feynmp-auto}
\usepackage{aas_macros}

\newcommand{\be}{\begin{eqnarray}}
\newcommand{\ee}{\end{eqnarray}}

\usepackage{soul}

\usepackage{lipsum}
\usepackage{orcidlink}
\definecolor{tabblue}{RGB}{31, 119, 180}
\definecolor{darkblue}{RGB}{0, 0, 120}
\definecolor{tabred}{RGB}{214, 39, 40}
\definecolor{tabgreen}{RGB}{44, 160, 44}
\definecolor{tabgray}{RGB}{100, 100, 100}
\definecolor{taborange}{RGB}{255, 127, 14}
\definecolor{tabbrown}{RGB}{128, 0, 0}
\definecolor{tabpink}{RGB}{255, 141, 161}
\definecolor{tabpurple}{RGB}{148, 103, 189}
\definecolor{goldenrod}{RGB}{218, 165, 32}
\usepackage{float}
\usepackage{placeins}
\newcounter{smsubcnt}
\makeatletter
\newcommand{\smtocsec}[2]{%
  \setcounter{smsubcnt}{0}%
  \par\addvspace{1.0em \@plus\p@}%
  \begingroup
    \hyphenpenalty=10000 \exhyphenpenalty=10000
    \sloppy
    \noindent
    \hangindent=2em\hangafter=1
    \hyperref[#2]{\makebox[2em][l]{\ref{#2}.}#1}%
    \nobreak\hfill\nobreak\hyperref[#2]{\pageref{#2}}\par
  \endgroup
}
\newcommand{\smtocsub}[2]{%
  \stepcounter{smsubcnt}%
  \begingroup
    \hyphenpenalty=10000 \exhyphenpenalty=10000
    \sloppy
    \noindent
    \hangindent=4em\hangafter=1
    \hspace*{2em}\hyperref[#2]{\makebox[2em][l]{\Alph{smsubcnt}.}#1}%
    \nobreak\hfill\nobreak\hyperref[#2]{\pageref{#2}}\par
  \endgroup
}
\makeatother
\begin{document}

\preprint{\hbox{UTWI-16-2026, FERMILAB-PUB-26-0307-T}}

\title{WIMP-like Dark Matter Without Thermalization At Freeze-Out}

\author{Dan Hooper\,\orcidlink{0009-0004-2456-1221}}
\email{dwhooper@wisc.edu}
\affiliation{Department of Physics, University of Wisconsin, Madison, WI, 53703, USA}
\affiliation{Wisconsin IceCube Particle Astrophysics Center (WIPAC),
University of Wisconsin, Madison, WI, 53703, USA}

\author{Gordan Krnjaic\,\orcidlink{0000-0001-7420-9577}}
\email{krnjaicg@uchicago.edu}
\affiliation{Fermi National Accelerator Laboratory, Batavia, IL, 60510, USA}
\affiliation{Kavli Institute for Cosmological Physics, University of Chicago, Chicago, IL, 60637, USA}
\affiliation{Department of Astronomy and Astrophysics, University of Chicago, Chicago, IL, 60637, USA}

\author{Gabriele Montefalcone\,\orcidlink{0000-0002-6794-9064}}
\email{montefalcone@utexas.edu}
\affiliation{Texas Center for Cosmology and Astroparticle Physics, Weinberg Institute for Theoretical Physics, Department of Physics, The University of Texas at Austin, Austin, TX 78712, USA}

\begin{abstract}

In the standard thermal relic scenario, dark matter remains in chemical equilibrium with the Standard Model radiation bath until freeze-out occurs at $T \sim m_X/20$, where $m_X$ is the dark matter mass. In this familiar class of models, the observed relic density is obtained for annihilation cross sections of order $\sigma v \sim 10^{-26}$ cm$^3$/s.  We show that comparable cross sections can naturally be realized in hidden-sector models in which the dark matter and Standard Model sectors decouple at a very high temperature, $T \gg m_X$, and subsequently evolve with separate thermal histories. Despite this decoupling, the two sectors have similar temperatures during freeze-out, leading to the usual thermal relic abundance. As a consequence, the coupling between the Standard Model and hidden sectors can be extremely small, potentially placing direct detection and collider signals far below foreseeable sensitivities.

\end{abstract}
  
\maketitle
 
\textbf{\textit{Introduction---}} In the Weakly Interacting Massive Particle (WIMP)
paradigm, dark matter (DM) particles with roughly weak-scale masses
and couplings are initially in chemical equilibrium with the Standard Model (SM) bath. When the temperature of the universe falls below
the WIMP’s mass, these interactions freeze out, yielding
a thermal relic abundance that is consistent with the measured density of DM. This outcome has often been referred to as the ``WIMP miracle''~\cite{Bertone:2016nfn}. 

In light of stringent constraints from direct detection experiments~\cite{LZ:2024zvo,XENON:2023cxc,PICO:2019vsc} and collider searches for DM~\cite{ATLAS:2017bfj,CMS:2018mgb,ATLAS:2023rvb,ATLAS:2021kxv,CMS:2012ucb,ATLAS:2011kno}, it has become increasingly well motivated to consider DM candidates that are part of a so-called hidden sector, which does not couple directly to the particle content of the SM. Instead of annihilating into SM particles, hidden sector DM candidates annihilate into other hidden sector particle species that later decay through small couplings into the SM~\cite{Pospelov:2007mp,Arkani-Hamed:2008hhe,Berlin:2016gtr,Berlin:2016vnh,Patt:2006fw,McDonald:1993ex,Falkowski:2011xh,Cline:2014dwa,Escudero:2016ksa,Escudero:2016tzx,Escudero:2017yia,Evans:2017kti,Evans:2019vxr,Hooper:2019xss,Kong:2025ccv,Koechler:2025ryv}. 

If the SM and hidden sectors were at similar temperatures during freeze-out, such DM candidates would yield an acceptable relic abundance if their annihilation cross section were near that of a conventional WIMP, $\sigma v \sim 10^{-26} \, {\rm cm}^3/{\rm s}$. However, if these two sectors were never in equilibrium, there would be no obvious reason to expect their temperatures to be comparable. Instead, their relative temperatures (or corresponding energy densities) could be set by initial conditions, such as those associated with the era of reheating following inflation. In such a scenario, the resulting thermal relic abundance could vary considerably from that of an canonical WIMP, providing little motivation to favor any particular regions of parameter space~\cite{Berlin:2016gtr}. For this reason, studies of hidden sector DM commonly require thermal equilibrium to be maintained between the SM and hidden sectors during DM freeze-out~\cite{Evans:2017kti,Hooper:2019xss,Hu:2025thq}. Satisfying this requirement imposes a lower limit on the coupling between these sectors and thereby implies a lower bound on the cross sections probed by direct detection experiments. 
    
In this {\it letter}, we challenge key elements of this logic and show that, even if the coupling between these sectors is too feeble to maintain equilibrium during freeze-out, it is natural to expect that equilibrium was achieved at much earlier times. 
This prior history predicts that the two sectors should have similar temperatures during freeze-out, which restores the usual expectations for the DM annihilation cross section and thermal relic abundance. 

In light of this, we argue here that it is plausible that the coupling between the SM and hidden sectors could be extremely small, leading to significant implications for direct detection and DM collider searches. The magnitude of this coupling is bounded only by the successful predictions of Big Bang nucleosynthesis (BBN) and allow for the possibility that the DM could scatter with nuclei at a rate that is many orders of magnitude below the sensitivity of existing experiments, or even those at the level of the so-called ``neutrino fog.''

\begin{figure}
    \centering
    \includegraphics[width=\linewidth]{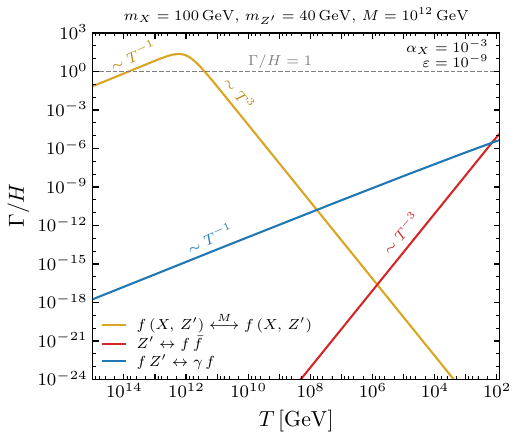}
    \caption{The ratio $\Gamma/H$ for several processes connecting the hidden and Standard Model (SM) sectors, shown as a function of the SM temperature, $T$. In gold, we show this ratio for SM--hidden-sector scattering processes that are mediated by a heavy particle of mass $M$. These processes decouple at $T \ll M$, but enforce equilibrium at $T\sim M$. In red and blue, respectively, we show the decay and scattering rates of the hidden-sector mediator, $Z^\prime$, with the SM bath. The rates for these processes are each controlled by the value of the portal coupling, $\varepsilon$. This benchmark model illustrates how  equilibrium between the two sectors could be maintained at high temperatures even when the portal coupling is too small to achieve equilibrium at lower temperatures.}
    \label{fig:1}
\end{figure}

\medskip

\textbf{\textit{Thermalization at High Temperatures---}} At the temperature of DM freeze-out, equilibrium between the SM and hidden sectors can be maintained through processes involving scattering, decays, and inverse decays. For relativistic particles in the thermal bath, scattering rates mediated by light particles typically scale as 
\be
\Gamma = n\, \langle \sigma v\rangle \propto  T^3  \, T^{-2} \sim T
,
\ee
where $n$ is the number density of targets in the bath. Comparing this to the Hubble rate during radiation domination, $H\propto T^2$, one finds
\be
\Gamma/H \propto T/T^2 \sim T^{-1}.
\ee
Therefore, once equilibrium is established, it is maintained as the universe expands and cools, at least until the particles in question become non-relativistic. 

However, this conclusion only follows if the interactions between these sectors are also mediated by a particle whose mass is light compared to the temperature. If we instead consider a process that is mediated by a very massive particle, $M \gg T$, the scattering rate scales as
\be
\Gamma = n \langle \sigma v\rangle  \propto T^3 (T^{2}/M^4) \propto T^5~ \implies ~ \Gamma/H \propto T^3\, ,
\ee
 and thus equilibrium can be maintained at very high temperatures but not necessarily during dark matter freeze-out. This requires only the existence of heavy particles with masses much greater than the temperature of DM freeze-out, which can mediate efficient scattering between the SM and hidden sectors at high temperatures. 

In a scenario in which equilibrium was reached between the two sectors at high temperatures, their temperatures remain approximately equal to each other after decoupling, differing only due to the entropy injection experienced as particles in each sector become non-relativistic. For this reason, we expect the temperature of the hidden sector to be similar to that of the SM bath at the time of DM freeze-out.

Heavy particles that would enable the two sectors to reach and maintain equilibrium at high temperatures are generically expected in many realistic UV completions of hidden sector DM models:
\begin{itemize}
    \item {\bf Hypercharge Portal.} 
    Consider, for example, the class of models in which SM particles interact with a hidden sector through the kinetic mixing of a hidden sector $U(1)$ with SM hypercharge, as parameterized by $\mathcal{L} \supset \frac{1}{2} \varepsilon B^{\mu \nu} Z'_{\mu \nu} $, where $B_{\mu \nu}$ and $Z^\prime_{\mu\nu}$ are the hypercharge and hidden field-strength tensors, respectively. Although any value of $\varepsilon$ is technically natural in the context of the effective theory~\cite{Arkani-Hamed:2008kxc}, if the SM is embedded in a Grand Unified Theory (GUT), a non-zero value of $\varepsilon$ will be generated after GUT breaking only through interactions that involve loops of heavy particles which carry both SM hypercharge and the hidden sector gauge charge. 
    
    At the one-loop level, these heavy states can be integrated out to induce the following degree of kinetic mixing~\cite{Holdom:1986eq}:
\be
~~~~~ ~~~~~\varepsilon \sim \frac{ g_X g_Y}{16 \pi^2} \ln \left(\frac{M'^2}{M^2}\right) 
~~~\text{(one-loop)},~
\ee
where $g_Y$ is the hypercharge coupling constant, $g_X$ is the hidden sector gauge coupling, and $M'/M$ is the ratio of the masses of the particles in the loop.  Alternatively, if this kinetic mixing is induced only at two-loop or higher order, we expect 
\be
~~~~~ ~~~~~ \varepsilon \sim  \frac{  g_X g_Y  \Pi_i^{2(n-1)} |y_i|^2  }{(16 \pi^2)^n} \ln \left(\frac{M'^2}{M^2}\right) ~~~~ \text{($n$-loop)},
\ee
where $n$ is the number of loops in those diagrams and we have assumed that scalars with  Yukawa couplings $y_i$ intersect fermion loops to generate these contributions.

\item {\bf Higgs Portal.} A Higgs portal coupling could also be radiatively generated through loops of heavy particles~\cite{DiFranzo:2015nli,DiFranzo:2016uzc,Voigt:2017vfz}. For example, if a heavy hidden-sector scalar $S$ has interactions of the form 
\be
~~~~~ ~~~~~ 
\mathcal{L} \supset -S( a |H|^2 - b |H'|^2) +\frac{ M^2_S}{2} S^2,
\ee
where $H$ and $H^\prime$ are visible and hidden-sector Higgs fields, 
then integrating out $S$ yields an effective portal interaction $\lambda|H^\prime|^2 |H|^2$,  where the coupling is  given by $\lambda \sim ab/M^2_{S}$~\cite{Dittmaier:2021fls,Frangipane:2021rtf,Bertuzzo:2024bwy}. 

\end{itemize}
In either case, the portal coupling arises from the presence of heavy particles that couple to both sectors.

In Fig.~\ref{fig:1}, we plot the ratio of several scattering and decay rates to the rate of Hubble expansion in the early universe as a function of temperature, for an example of a hidden sector DM model involving the hypercharge portal. Due to the very small value of the portal coupling, $\varepsilon =10^{-9}$, the rates for the processes $f Z' \leftrightarrow \gamma f$ and $Z' \leftrightarrow f\bar{f}$ (where $f$ is a SM fermion and $Z'$ is a hidden sector mediator) are each well below that of Hubble expansion, and are thus unable to maintain equilibrium between the SM and hidden sectors. In contrast, scattering processes mediated by a very heavy ($M =10^{12} \, {\rm GeV}$) state with both SM hypercharge and hidden sector gauge charge are sufficient to equilibrate the two sectors at early times. This behavior is a generic feature of hidden-sector scenarios with small portal couplings and heavy states charged under both sectors.

\medskip
\textbf{\textit{Hidden Sector Freeze-Out \& Entropy Dilution---}} 
We consider a hidden sector that includes a DM particle, $X$, which annihilates into unstable mediators, $Y$, through hidden-sector interactions, $X\bar{X} \to YY$. The mediators then decay into SM final states through a small portal coupling, $\varepsilon$. The cosmological evolution of this system is governed by the Boltzmann equations for the number densities of these particles, $n_X$ and $n_Y$, together with the conservation of energy in the hidden sector:
\be
    \dot{n}_X+3Hn_X &=&-\langle \sigma v\rangle_X\left[n_X^2-\frac{n_Y^2}{(n_Y^{\rm eq})^2}(n_X^{\rm eq})^2\right]\,, \label{eq:boltz_nX} \\
    \dot{n}_Y+3Hn_Y &=& +\langle \sigma v\rangle_X\left[n_X^2-\frac{n_Y^2}{(n_Y^{\rm eq})^2}(n_X^{\rm eq})^2\right] + C_{\rm dec}~,~~~~~
\label{eq:boltz_nY}
    \ee
    where $\langle \sigma v\rangle_X$ is the thermally averaged cross section for $X\bar{X}\to YY$ and 
     the collision term for decays is given by
     \be
C_{\rm dec} =   -\Gamma_{\rm can}\,(n_Y-n_Y^{\rm eq})-\Gamma_Y\, (n_Y - n_Y^{\rm eq, SM}),
     \ee     
where $\Gamma_Y \propto \varepsilon^2$ is the mediator decay rate into SM particles,  and $n_Y^{\rm eq, SM}$ denotes the mediator equilibrium density evaluated at the SM temperature, $T$ (entering through inverse decays). The cannibal reaction rate is given by
\begin{equation}
    \Gamma_{\rm can} = \langle \sigma v^2\rangle_{YX}\, n_Y\, n_X + \langle \sigma v^2\rangle_{X}\, n_X^2 + \langle \sigma v^2\rangle_{Y}\, n_Y^2,
    \label{eq:Gamma_can}
\end{equation}
which accounts for $3\to 2$ number-changing processes within the hidden sector, where $\langle \sigma v^2\rangle_{YX}$, $\langle \sigma v^2\rangle_{X}$ and $\langle \sigma v^2\rangle_{Y}$ denote the thermally averaged cross sections for $YYX\rightarrow YX$, $YXX\rightarrow XX$ and $YYY\rightarrow YY$, respectively.  
     The hidden sector energy density satisfies
     \be
    \dot{\rho}_h+3H(\rho_h+P_h) &=&-\Gamma_Y\,(\rho_Y -\rho^{\rm eq, SM}_Y)\,,  \label{eq:boltz_rho}
\ee
where density and pressure are $\rho_h = \rho_X+\rho_Y$ and $P_h = P_X+P_Y$, and the Hubble rate, $H^2 = 8\pi \,(\rho_{\rm SM}+\rho_h)/3M_{\rm Pl}^2$, includes contributions from both sectors, with $M_{\rm Pl} \approx 1.22\times 10^{19}\,$GeV.

All other equilibrium densities, $n_i^{\rm eq}$, are evaluated at the hidden-sector temperature, $T_h$,
which evolves independently of the SM temperature, $T$, when the portal coupling is too feeble to maintain thermal equilibrium between the two sectors. The ratio of these temperatures is determined by entropy conservation;
\begin{align}
\xi \equiv \frac{T_h}{T} = \left[
\frac{g_{\star, s}(T)}{g_{\star, s}(T_{\rm ini})}\,
\frac{g^h_{\star, s}(T_{\rm ini})}{g^h_{\star, s}(T_h)}\right]^{1/3},
\end{align}
where $g_{\star, s}$ and $g_{\star, s}^h$ are the effective number of entropic degrees of freedom in the SM and hidden sectors, respectively, and $T_{\rm ini}$ is an initial temperature at which the two sectors were in equilibrium, $\xi =1$. Unless there are many particle species with masses between $T_{\rm ini}$ and $T \sim {\rm TeV}$, we can expect $\xi \approx 1$ at $T\sim {\rm TeV}$ (where our Boltzmann code begins). With this in mind, we take $\xi_{\rm ini} \approx 1$ as our initial condition. If $g_{\star, s}(T_{\rm ini})$ or $g_{\star, s}^h(T_{\rm ini})$ were very large, however, the value of $\xi$ at $T \sim {\rm TeV}$ could depart significantly from unity.

The cosmological evolution of this system naturally separates into two stages:
\begin{enumerate}
    \item {\bf Secluded Evolution.} First, the hidden sector undergoes internal chemical freeze-out. The hidden-to-SM temperature ratio, $\xi$, evolves through entropy conservation in each sector, possibly modified by a transient cannibal phase once the mediator becomes non-relativistic~\cite{Berlin:2016vnh}, but remains of order unity throughout. Since $\Gamma_Y \ll H$ during this entire epoch, the portal coupling, $\varepsilon$, plays no role and the resulting abundances are determined almost entirely by the hidden-sector masses and couplings.

    \item {\bf Entropy Transfer.} In the second stage, the frozen-out mediator population, redshifting as matter, comes to dominate the total energy density before decaying into SM particles. These decays inject entropy into the visible sector, diluting all pre-existing comoving abundances by a factor of~\cite{Berlin:2016vnh, Berlin:2016gtr}
\begin{equation}
    \frac{S_{f}}{S_{i}}  = \frac{~\rho_Y(\tau_Y)^{3/4}}{T(\tau_Y)^3},
\end{equation} 
where $S_{i,f}$ are the SM entropies before and after $Y$ decay, respectively, and we have used the sudden-decay approximation. 
Since $\Gamma_Y \propto \varepsilon^2$, smaller portal couplings correspond to longer mediator lifetimes, greater entropy injection, and thus to stronger dilution of the relic abundance. For a given hidden-sector coupling, this allows the relic abundance to match the measured DM density for a wide range of DM masses, including those well above the canonical thermal relic value, 
while tying the predicted density to the same portal coupling, $\varepsilon$, that governs the mediator lifetime and the DM scattering cross section with nuclei.
\end{enumerate}

The role of entropy dilution in enabling heavy DM candidates has long been appreciated~\cite{Berlin:2016vnh, Berlin:2016gtr} (see also, Refs.~\cite{Fornengo:2002db,Gelmini:2006pq,Kane:2015jia,Hooper:2013nia,Patwardhan:2015kga,Randall:2015xza,Patwardhan:2015kga,Reece:2015lch,Lyth:1995ka,Davoudiasl:2015vba,Cohen:2008nb,Boeckel:2011yj,Boeckel:2009ej,Hong:2015oqa}). In this {\it letter}, we emphasize a complementary implication of this mechanism: since hidden freeze-out proceeds largely as it would for a standard thermal relic --- with $\xi$ remaining of order unity throughout --- the annihilation cross section required of a hidden sector relic is comparable to that of a standard thermal relic, even for values of $\varepsilon$ that are far too small to maintain equilibrium between the two sectors at the time of freeze-out.  Furthermore, previous analyses of this mechanism have largely focused on the $m_X \gg m_Y$ regime, in which DM freezes out while the mediator remains relativistic and in chemical equilibrium, thereby significantly simplifying the evolution~\cite{Berlin:2016vnh, Berlin:2016gtr}. Here, however, we extend the analysis to mass ratios of order a few, for which both species are non-relativistic at the time of freeze-out.  In this regime, the full coupled system of Eqs.~\eqref{eq:boltz_nX}--\eqref{eq:boltz_rho}, including the cannibal processes of Eq.~\eqref{eq:Gamma_can}, must be solved numerically. For further details, we direct the reader to the Supplementary Material.

\begin{figure}
    \centering
    \includegraphics[width=\linewidth]{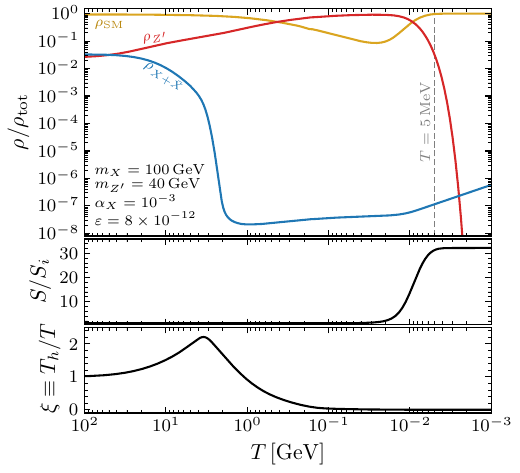}
    \caption{The cosmological evolution of a hypercharge portal dark matter model for a representative choice of model parameters, as a function of the temperature of the Standard Model (SM) bath, $T$. In the {\it top panel}, we show the normalized energy densities of dark matter, $X+\bar X$, the hidden-sector mediator, $Z'$, and the SM bath. This illustrates how the $Z'$ population comes to dominate the energy density after hidden-sector freeze-out, and then decays to reheat the SM bath. The {\it  middle panel} quantifies the resulting dilution of pre-existing comoving abundances through the comoving entropy ratio, $S/S_i$. The {\it bottom panel} shows the hidden-to-SM temperature ratio, $\xi = T_h/T$ , which remains of order unity throughout freeze-out and subsequently tracks its asymptotic non-relativistic scaling.}
    \label{fig:2}
\end{figure}

\medskip

\textbf{\textit{A Concrete Example: The Hypercharge Portal---}} The mechanism described above can be realized within the context of several portal interactions connecting the hidden sector to the SM. In this {\it letter}, we focus on the hypercharge portal as a representative example,  with additional portals discussed in the Supplementary Material. The relevant interactions are described by the following Lagrangian:
\begin{equation}
    \mathcal{L}\supset  -\frac{\varepsilon}{2}B^{\mu\nu}Z^\prime_{\mu\nu}+g_XZ^\prime_\mu\bar{X}\gamma^\mu X,
\end{equation}
where $B^{\mu\nu}$ and $Z'_{\mu\nu}$ are the hypercharge and dark $U'(1)$ field strength tensors, $g_X$ is the dark gauge coupling, and $\varepsilon$ is the kinetic mixing parameter introduced earlier. Our DM candidate, $X$, is a stable Dirac fermion charged under the spontaneously broken dark $U^\prime(1)$ gauge group, and $Z^\prime$ is the gauge boson associated with this symmetry. The $Z'$ will play the role of the hidden sector mediator, $Y$. 

Hidden-sector freeze-out is governed by $X\bar{X}\to Z'Z'$ annihilation, whose thermally averaged cross section in the low-velocity limit is given by
\begin{equation}
    \langle\sigma v\rangle_X \approx \frac{4\pi\alpha_X^2}{m_X^2}\,\frac{(1-r^2)^{3/2}}{(2-r^2)^2}\,,
    \label{eq:sv_swave}
\end{equation}
where $\alpha_X\equiv g_X^2/(4\pi)$ and $r\equiv m_{Z^\prime}/m_X$. Note that we show here only the leading velocity-independent contribution while the full expression, together with the thermally averaged cross sections for the $Z^\prime$ cannibal processes, can be found in the Supplementary Material. The kinetic mixing parameter, $\varepsilon$, enters the cosmological evolution only through the $Z^\prime$ decay width, controlling the entropy dilution and impacting the final DM relic abundance as described above. 

The two stages of the cosmological evolution can be clearly identified in Fig.~\ref{fig:2} for a representative choice of model parameters. In the {\it top panel}, we plot the energy densities of the visible and hidden sectors, showing how the $Z^\prime$ population redshifts as non-relativistic matter after freeze-out and eventually comes to dominate the total energy density before decaying into SM particles. These decays reheat the SM bath and dilute all pre-existing comoving abundances, as shown by the entropy ratio in the {\it middle panel}, which for these parameters plateaus at $S_{f}/S_{i} \sim 30$. In the {\it bottom panel}, we show the evolution of the hidden-to-SM temperature ratio, $\xi$, which offers a direct view of the freeze-out dynamics. Starting from $\xi = 1$ --- reflecting early-time thermalization between the two sectors (see Fig.~\ref{fig:1}) --- $\xi$ increases through a transient cannibal phase as the $Z'$ population becomes non-relativistic while still remaining in chemical equilibrium, and subsequently tracks the non-relativistic scaling, $T_h \propto 1/a^{2}$, or $\xi\propto1/a$, once the hidden sector freezes out.

\begin{figure*}
    \centering
    \includegraphics[width=0.9\linewidth]{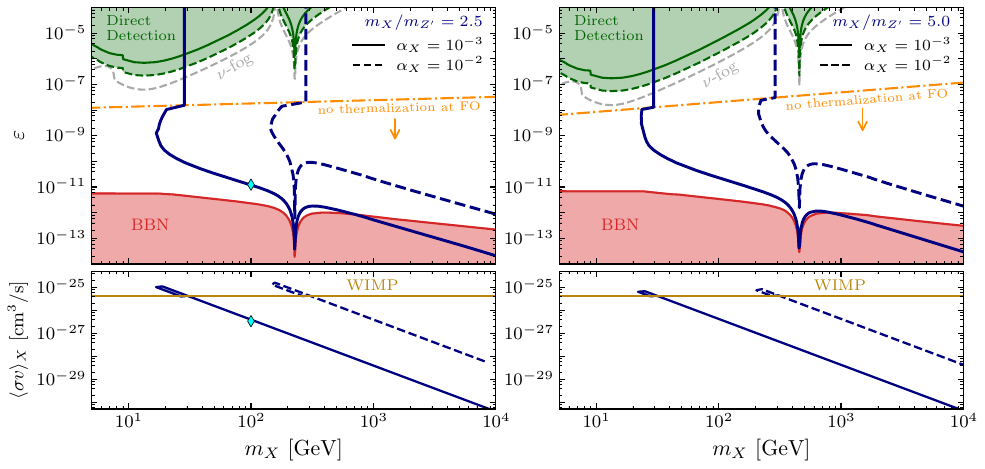}
    \caption{Viable parameter space in the $(m_X,\varepsilon)$ plane for the hypercharge portal model, for mass ratios of $m_X/m_{Z'} = 2.5$~({\it left}) and $m_X/m_{Z'} = 5$~({\it right}), and with $\alpha_X = 10^{-3}$ (solid) and $\alpha_X = 10^{-2}$ (dashed). In the {\it top panels}, the blue curves denote the value of $\varepsilon$ that is required to produce the observed dark matter relic abundance, $\Omega_X h^2 = 0.12$. The red and green shaded regions correspond, respectively, to the constraints derived from the primordial light element abundances and direct-detection experiments, specifically LZ~\cite{LZ:2024zvo} and XENONnT~\cite{XENON:2025vwd}, while the dot-dashed orange line marks the threshold above which thermal equilibrium between the two sectors is maintained until freeze-out (FO). The dashed gray curve shows the (more conservative) neutrino fog for $\alpha_X = 10^{-2}$. The cyan diamond marks the benchmark of Fig.~\ref{fig:2}. The {\it bottom panels} show the corresponding low-velocity DM annihilation cross section, $\langle \sigma v\rangle_{X}$, evaluated along the DM relic curves, compared to the value required for a thermalized WIMP (dot-dashed gold). Overall, the figure shows that the observed relic abundance can be reproduced for portal couplings far below current direct-detection sensitivity, while maintaining an annihilation cross section close to the thermal-WIMP value. 
    }
    \label{fig:3}
\end{figure*}

\medskip

\textbf{\textit{Observational Constraints---}} In the {\it top panels} of Fig.~\ref{fig:3}, we show the constraints on the hypercharge portal parameter space for two values of the hidden sector coupling, $\alpha_X$, and the mass ratio, $m_X/m_{Z'}$. These panels include curves along which the predicted relic abundance matches the measured DM density, $\Omega_X h^2 = 0.12$~\cite{planck18}. Large values of the portal coupling, $\varepsilon$, are constrained by direct detection experiments~\cite{LZ:2024zvo,XENON:2025vwd}, while very small values are excluded by measurements of the primordial element abundances (for further details, see the Supplementary Material). 

In the {\it bottom panels} of Fig.~\ref{fig:3}, we plot the DM annihilation cross section evaluated at low-velocity, as is typically relevant for indirect searches. This is compared to the value predicted for a conventional WIMP with a velocity-independent annihilation cross section, $\langle \sigma v \rangle_{\rm WIMP} \approx 4.4 \times 10^{-26} \, {\rm cm}^3/{\rm s}$ (for a Dirac fermion). For values of the portal coupling $\varepsilon \gtrsim 10^{-10}$, the decay of the $Z'$ population does not significantly dilute the relic abundance, and a WIMP-like annihilation cross section is required to produce the observed DM density. For smaller values of $\varepsilon$, the entropy injection dilutes the DM abundance, thus requiring a smaller value of $\langle \sigma v \rangle$.

\medskip

\textbf{\textit{Discussion---}} It is commonly believed that if dark matter (DM) is not in equilibrium with the Standard Model (SM) during freeze-out, there is little reason to expect its thermal-relic abundance to match the observed DM density, and correspondingly little motivation for any particular value of the DM annihilation cross section. With this lore in mind, some studies of hidden sector DM explicitly demand that model parameters be compatible with thermalization during DM freeze-out \cite{Evans:2017kti,Hooper:2019xss,Hu:2025thq}. This requirement, in turn, implies a lower limit on the predicted DM-SM scattering cross section, which serves as an important milestone for direct detection searches.
    
In this {\it letter}, we show that even if the couplings between the hidden and SM sectors are too feeble to maintain equilibrium at the time of freeze-out, equilibrium could reasonably be achieved at much earlier times with only minimal assumptions about high-energy physics. If the DM and SM sectors decouple while both are still relativistic, they naturally  maintain similar temperatures during freeze-out, so the DM still acquires a thermal-relic abundance with WIMP-like properties. Given the simplicity of this scenario, is plausible that the coupling between the SM and hidden sectors could be too feeble for direct detection and collider searches.

Nonetheless, these hidden sector scenarios with tiny portal couplings and WIMP-like annihilation cross sections could potentially lie within the reach of current or future indirect searches. In future work, we plan to explore this class of models, particularly in the context of the Galactic Center Gamma-Ray Excess~\cite{Goodenough:2009gk,Hooper:2010mq,Hooper:2011ti,Abazajian:2012pn,Hooper:2013rwa,Gordon:2013vta,Daylan:2014rsa,Calore:2014xka,Fermi-LAT:2015sau,Fermi-LAT:2017opo,Cholis:2021rpp,DiMauro:2021raz}.

\medskip

\textbf{\textit{Acknowledgments---}} GM would like to thank Barmak Shams Es Haghi for helpful discussion. DH is supported by the Office of the Vice Chancellor for Research at the University of Wisconsin-Madison, with funding from the Wisconsin Alumni Research Foundation. GM acknowledges support by the Writing Fellowship of the Graduate School of the College of Natural Sciences at the University of Texas at Austin. 
This manuscript has been authored in part by Fermi Forward Discovery Group, LLC under Contract No. 89243024CSC000002 with the U.S. Department of Energy, Office of Science, Office of High Energy Physics. 
We acknowledge the use of the Mathematica package \texttt{FeynCalc}~\cite{Mertig:1990an, Shtabovenko:2016sxi, Shtabovenko:2020gxv, Shtabovenko:2023idz} for the computation of the hidden-sector cross sections, and of the \texttt{HDECAY}~\cite{Djouadi:1997yw,Djouadi:2018xqq} and \texttt{scalar\_portal}~\cite{Winkler:2018qyg} packages for the evaluation of the scalar mediator decay width in the Higgs portal.

\bibliographystyle{apsrev4-1}
\bibliography{bibl}
\clearpage
	\onecolumngrid
\begin{center}
  \textbf{\large Supplementary Material for WIMP-like Dark Matter Without Thermalization At Freeze-Out}\\[.2cm]
  \vspace{0.05in}
  {Dan Hooper, Gordan Krnjaic, and Gabriele Montefalcone}
\end{center}
\vspace{1 cm}
	\twocolumngrid
\section*{Contents}

\renewcommand{\theequation}{S\arabic{equation}}
\setcounter{equation}{0}

\renewcommand{\thefigure}{S\arabic{figure}}
\setcounter{figure}{0}

\smtocsec{Details of the Numerical Implementation}{sm:num}
\smtocsub{Boltzmann Equations}{sm:boltz}
\smtocsub{Three-Phase Solver Strategy}{sm:solver}
\smtocsub{Late-Time Evolution \& Mediator Decay}{sm:latetime}
\smtocsub{Approaching the Thermalization Floor}{sm:thermfloor}

\smtocsec{Hypercharge Portal: Cross Sections and Rates}{sm:hyper}
\smtocsub{Hidden-sector cross sections}{sm:hyper_dark}
\smtocsub{Mediator decay width}{sm:hyper_decay}
\smtocsub{Spin-independent DM--nucleon cross section}{sm:hyper_si}

\smtocsec{Constraints from Big Bang Nucleosynthesis}{sm:bbn}

\smtocsec{Results for Other Portals}{sm:other}
\smtocsub{The Higgs Portal}{sm:higgs}
\smtocsub{The $B-L$ Portal}{sm:bml}
\smtocsub{The Baryon Portal}{sm:baryon}

\section{Details of the Numerical Implementation}\label{sm:num}

\noindent In this section, we provide a concise but complete description of the numerical framework used to compute the hidden-sector dynamics, along with the key assumptions underlying our treatment. A public implementation of the full \texttt{hidden\_sector\_DM} toolkit is available  on \texttt{github}~\href{https://github.com/GabrieleMonte/hidden_sector_DM/}{\faGithub}. 
Before turning to the equations and their numerical solution, we spell out the three main assumptions that underlie our treatment. 

First, we model all hidden-sector species using Maxwell-Boltzmann (MB) statistics. The deviation of this approximation from the full quantum distributions is largest in the relativistic regime, $m/T_h \lesssim 1$, but becomes exponentially suppressed by the time of freeze-out, $m/T_h \gtrsim 20$–$30$. As a result, the net impact of this on the computed relic abundance is well below the percent level.

Second, we assume that the hidden sector remains in internal kinetic equilibrium at a common temperature, $T_h$, throughout. This is justified by the fact that elastic scattering $XY\to XY$ shares the same $\alpha_X^2$ scaling as the $2\to 2$ DM annihilation but is enhanced by a factor of $n_Y/n_X\gg 1$ throughout the freeze-out epoch, since the lighter mediator is less Boltzmann-suppressed than the DM. Elastic equilibration is therefore always the fastest process in the hidden sector, justifying the assignment of a single common temperature throughout~(see also Ref.~\cite{Fernandez:2021iti}). 

Third, we treat the portal coupling, $\varepsilon$, as decoupled from the dynamics of freeze-out dynamics. This is justified since the decay rate $\Gamma_Y \propto \varepsilon^2$ is negligible compared to the Hubble rate throughout the Boltzmann evolution. We therefore solve the Boltzmann system with $\Gamma_Y = 0$, reintroducing the decay only in the late-time evolution that follows freeze-out. This two-step procedure agrees with a fully joint evolution at the sub-percent level for $\varepsilon \lesssim 10^{-10}$, which comfortably covers the majority of the parameter space relevant to this work.

\subsection{Boltzmann Equations}\label{sm:boltz}
For each species $i$, the Boltzmann system can be written
\begin{align}
    &\rho_i = B_1^i(T_h)\,n_i = m_i\,\left[\frac{K_1(m_i/T_h)}{K_2(m_i/T_h)} + \frac{3\,T_h}{m_i}\right]\,n_i\,, \\
    &\rho_i + P_i = B_2^i(T_h)\,n_i = \left[B_1^i(T_h) + T_h\right]\,n_i\,,
\end{align}
where the index $i\in \{X,\,Y\}$ runs over the hidden-sector species, $K_1$ and $K_2$ are modified Bessel functions of the second kind, and the coefficients $B_1^i(T_h)$ and $B_2^i(T_h)$ encode the average energy and enthalpy per particle, respectively. We emphasize that these relations hold for the actual $n_i$ throughout the freeze-out process, even as the species fall out of chemical equilibrium, since the kinetic-equilibrium assumption stated above is sufficient to preserve the MB form of the distribution function. Substituting these expressions into Eq.~\eqref{eq:boltz_rho}, the hidden-sector energy conservation equation reduces to an evolution equation for $T_h$ given $n_X$ and $n_Y$. 

For numerical convenience, we replace $T_h$ with the hidden-to-SM temperature ratio, $\xi \equiv T_h/T$, which is expected to start from $\xi_{\rm ini} = 1$ given the high-temperature thermalization between the two sectors discussed in the main text, and evolves only mildly thereafter. In parallel, rather than evolving the number densities $n_X$ and $n_Y$ directly, we reformulate the system in terms of the dimensionless chemical potentials, $\bar\mu_i \equiv \mu_i/T_h$, defined through
\begin{equation}
    Y_i = Y_i^{\rm eq}\,e^{\bar\mu_i}\,, \qquad i \in \{X,Y\}\,,
    \label{eq:chempot_def}
\end{equation}
where $Y_i = n_i/s$ is the comoving yield, $s\equiv\left(  2\pi^2/45\right)\,g_{\star, s}(T)\,T^3$ is the SM entropy density, $g_{\star,s}$ is the effective number of relativistic degrees of freedom in entropy, $Y_i^{\rm eq} = n_i^{\rm eq}/s$, and $n_i^{\rm eq}(T_h) = g_i\, m_i^2\, T_h\,K_2(m_i/T_h)/(2\pi^2)$
is the MB equilibrium number density of the $i$th species with $g_i$ internal degrees of freedom. 

Thus, in terms of $(\bar\mu_X,\,\bar\mu_Y,\,\xi)$, we obtain a coupled system of three ordinary differential equations (ODEs) in the dimensionless time variable, $x\equiv m_X/T$:
\begin{align}
    \frac{d\bar\mu_X}{dx} &= A + B\,\frac{d\ln\xi}{dx}\,, \label{eq:dmuX} \\
    \frac{d\bar\mu_Y}{dx} &= C + D\,\frac{d\ln\xi}{dx}\,, \label{eq:dmuY} \\
    \frac{d\ln\xi}{dx}    &= \frac{E + F\,A + G\,C}{1 - F\,B - G\,D}\,, \label{eq:dlnxi}
\end{align}
where the six coefficient functions $A-G$ depend on $(x, \,\xi,\, \bar\mu_X,\, \bar\mu_Y)$  and are constructed from the thermally averaged cross sections, the MB equilibrium thermodynamic factors $(B_1^i,\,B_2^i)$, and the SM entropy-conservation factor, $\tilde g\equiv 1+\left(d \ln g_{\star,S}/d\ln T\right)/3$. The DM coefficients read
\begin{align}
    A =& -\frac{s\,\tilde g}{H_{\rm tot}\,x}\,\langle\sigma v\rangle_X\,Y_X^{\rm eq}\,e^{\bar\mu_X}
         \left[1 - e^{2(\bar\mu_Y - \bar\mu_X)}\right] \nonumber \\
         &+ \frac{\lambda_X - 3\tilde g}{x}\,, \label{eq:Acoeff} \\
    B =& -\lambda_X\,, \label{eq:Bcoeff}
\end{align}
where the first term in $A$ is the collision integral for $X \bar X \leftrightarrow YY$, written in a form that makes the approach to chemical equilibrium ($\bar\mu_X = \bar\mu_Y)$ manifest, and the second encodes the redshifting of the equilibrium yield through $\lambda_X \equiv d\ln n_X^{\rm eq}/d\ln T_h= m_X^2 B_1^X/T_h$ and the entropy-running factor, $\tilde g$. The mediator coefficients take an analogous form,
\begin{align}
        C =& \frac{\tilde g}{H_{\rm tot}\,x}\,\Bigg\{
           s\,\langle\sigma v\rangle_X\,\frac{(Y_X^{\rm eq})^2}{Y_Y^{\rm eq}}\,e^{2\bar\mu_X - \bar\mu_Y}\,\left[1 - e^{2(\bar\mu_Y - \bar\mu_X)}\right]
           \nonumber \\
           &- \Gamma_{\rm can}\,\left(1 - e^{-\bar\mu_Y}\right)\,\Bigg\}
         + \frac{\lambda_Y - 3\tilde g}{x}\,, \label{eq:Ccoeff} \\
    D =& -\lambda_Y\,, \label{eq:Dcoeff}
\end{align}
where the first contribution in $C$ is the annihilation source, equal and opposite to the corresponding term in $A$. The second contribution involves the total cannibal rate, $\Gamma_{\rm can}$, defined in Eq.~\ref{eq:Gamma_can}, multiplied by the factor $(1 - e^{-\bar\mu_Y})$. This term vanishes identically when $Y$ is in chemical equilibrium ($\bar\mu_Y = 0$) and acts as a restoring force that drives $\bar\mu_Y$ back toward zero whenever cannibal reactions are fast. The last term plays the same role as in $A$, accounting for the redshifting of $Y_Y^{\rm eq}$. Finally, the hidden-sector temperature-equation coefficients arise from projecting the energy-conservation equation onto $d\ln\xi/dx$ and read
\begin{align}
    E &= \frac{1}{x}\,\left(1 - 3\tilde g\,\frac{n_X\,B_2^X + n_Y\,B_2^Y}{\mathcal{D}}\right), \label{eq:Ecoeff}\\
    F &= -\frac{B_1^X\,n_X}{\mathcal{D}}\,, \label{eq:Fcoeff}\\
    G &= -\frac{B_1^Y\,n_Y}{\mathcal{D}}\,, \label{eq:Gcoeff}
\end{align}
where the energy denominator is given by
\begin{align}
    \mathcal{D} =& T_h\,\left(n_X\,\frac{dB_1^X}{dT_h} + n_Y\,\frac{dB_1^Y}{dT_h}\right)  \nonumber \\
    &+ B_1^X\,n_X\,\lambda_X + B_1^Y\,n_Y\,\lambda_Y\,. \label{eq:Dtilde}
\end{align}
Here, $E$ captures the adiabatic cooling of the hidden sector due to Hubble expansion, while $F$ and $G$ encode the feedback of changes in the number densities on $T_h$, all controlled by the denominator, $\mathcal{D}$.

Two features of this parametrization make it considerably better suited to numerical integration than the direct evolution of the number densities. First, the source terms in $A$ and $C$ carry factors of $1 - e^{2(\bar\mu_Y - \bar\mu_X)}$ and $1 - e^{-\bar\mu_Y}$ that vanish identically in chemical equilibrium ($\bar\mu_i = 0$), bypassing the delicate numerical cancellation between $n_i^2$ and $(n_i^{\rm eq})^2\,(n_j/n_j^{\rm eq})^2$ that plagues the standard formulation when both are large and nearly equal. Second, near freeze-out the comoving yield, $Y_i$, departs exponentially from its equilibrium value, while $\bar\mu_i$ grows only linearly in $x$, making the dimensionless chemical potentials better behaved as ODE variables.

A final comment on conventions. The Boltzmann equations of the main text are written for a self-conjugate $X$ (Majorana fermion), following Ref.~\cite{Berlin:2016gtr}. For a Dirac $X$, $n_X$ counts both particles and antiparticles, and the cross sections $\langle\sigma v\rangle_X$ and $\langle\sigma v^2\rangle_{X}$ are each replaced by half their value to avoid double-counting identical pairs in the initial state.

\subsection{Three-Phase Solver Strategy}
\label{sm:solver}

\noindent  The system of Eqs.~\eqref{eq:dmuX}--\eqref{eq:dlnxi} is solved by exploiting the natural hierarchy of timescales in the hidden sector. The DM is the first species to depart from chemical equilibrium, at a time when the cannibal reactions are still fast enough to hold the mediator at its equilibrium abundance. Only later do the cannibal rates themselves become inefficient, and the mediator follows the DM out of equilibrium, entering the full freeze-out regime. Solving the complete three-variable system from the outset would introduce unnecessary stiffness in the early evolution, where the mediator equation is effectively algebraic. We therefore decompose the integration into three phases, progressively activating degrees of freedom as each process falls out of equilibrium.

In Phase~1, both species are in chemical equilibrium and $\bar\mu_X = \bar\mu_Y = 0$. The system reduces to a single ODE for $\ln\xi$, obtained from Eq.~\eqref{eq:dlnxi} with all equilibrium densities evaluated at $T_h$. The solver remains in this phase until the dimensionless DM annihilation rate, $\Gamma_{\rm ann}/H \equiv (s/H_{\rm tot})\,\langle\sigma v\rangle_X\,Y_X$, drops below a fixed threshold large enough to ensure that $X$ has not begun to depart substantially from equilibrium yet.

Phase~1.5 is the key ingredient that resolves the stiffness associated with the annihilation freeze-out transition. Here the DM has fallen out of chemical equilibrium, but the cannibal reactions remain fast enough to enforce $\bar\mu_Y = 0$. Setting $\bar\mu_Y = 0$ and $d\bar\mu_Y/dx = 0$ in Eqs.~\eqref{eq:dmuX}--\eqref{eq:dlnxi} eliminates the mediator equation altogether, reducing the system to two ODEs for $(\ln\xi,\, \bar\mu_X)$ with the mediator abundance fixed at its equilibrium value. This phase ends once the cannibal rate, $\Gamma_{\rm can}$, falls below the Hubble rate, beyond which the cannibal reactions can no longer maintain $Y$ in chemical equilibrium.

In Phase~2, all three variables $(\ln\xi,\, \bar\mu_X,\, \bar\mu_Y)$ evolve according to the full system of Eqs.~\eqref{eq:dmuX}--\eqref{eq:dlnxi} with the coefficients of Eqs.~\eqref{eq:Acoeff}--\eqref{eq:Dtilde}. Integration continues until the system has converged onto the asymptotic non-relativistic regime, in which the hidden-sector temperature follows $d\ln\xi/d\ln x = 1 - 2\tilde g$, the scaling expected once all species have frozen out and redshift as cold matter relative to the SM bath.

The advantage of this decomposition is twofold. First, the lower-dimensional Phases~1 and~1.5 avoid the stiffness that would otherwise arise from numerically enforcing chemical equilibrium through near-cancellations rather than imposing it by construction. Second, each $\bar\mu_i$ enters the dynamics only when its associated species is genuinely out of equilibrium, where it is well-behaved as an ODE variable. The trade-off is the need for transition criteria between phases, which we set conservatively to ensure the system has fully entered the relevant regime before each new variable is activated.

\subsection{Late-Time  Evolution \& Mediator Decay}
\label{sm:latetime}

\noindent  Once the Boltzmann system has converged, the comoving yields of $X$ and $Y$ are effectively frozen and the hidden-sector temperature has settled onto the asymptotic non-relativistic scaling. The subsequent evolution is governed by the decay of the mediator into SM particles, $Y \to \mathrm{SM}$, with the width $\Gamma_Y \propto \varepsilon^2$ now becoming the only relevant timescale and providing the only handle through which the portal coupling enters the predicted relic abundance. We evolve the system in the number of e-folds, $N_e = \ln(a/a_0)$, where $a_0$ is the scale factor at which the Boltzmann system was terminated, tracking the SM temperature, $T$, and the energy densities, $\rho_X$, $\rho_Y$, through
\begin{align}
    \frac{d\ln T}{dN_e} &= -\frac{1}{\tilde g}\,\left(1 - \frac{\rho_Y\,\Gamma_Y}{3\,H \,s\,T}\right)\,, \label{eq:dlnTdN} \\
    \frac{d\ln\rho_Y}{dN_e} &= -\left(3 + \frac{\Gamma_Y}{H}\right)\,, \label{eq:dlnrhoYdN} \\
    \frac{d\ln\rho_X}{dN_e} &= -3 - \frac{\langle\sigma v\rangle_{\rm s\text{-}wave}\,\rho_X}{m_X\,H }\left[1-\left(\frac{n_Y\, n_X^{\rm eq}}{n_X\,n_Y^{\rm eq}}\right)^2\right]\,, \label{eq:dlnrhoXdN}
\end{align}
where the Hubble rate is evaluated at each step from the Friedmann equation, $H^2=8 \pi \left(\rho_h +\rho_{\rm SM}\right)/(3M_{\rm Pl}^2)$, and $\langle \sigma v\rangle_{\rm s\text{-}wave}$ is the velocity-independent piece of the DM annihilation cross section, which fully captures the residual annihilation rate at this stage given the non-relativistic motion of the DM after freeze-out. Eq.~\eqref{eq:dlnTdN} captures the SM temperature evolution with an entropy-injection term from $Y \to \mathrm{SM}$ that slows the standard $T \propto a^{-1}$ redshifting. Eq.~\eqref{eq:dlnrhoYdN} combines the Hubble dilution of $\rho_Y$ with its depletion by decay, while Eq.~\eqref{eq:dlnrhoXdN} tracks $\rho_X$, retaining the residual contributions from both annihilations and inverse annihilations, which are typically negligible at this stage but are retained for completeness. Since the hidden sector is deep in the non-relativistic regime during this epoch, we use $n_j = \rho_j/m_j$ for $j\in\{X,Y\}$, and evaluate the equilibrium densities at the redshifted hidden-sector temperature, $T_h \propto a^{-1}$, appropriate for frozen-out matter species.

The relevant observable is the asymptotic DM yield, $Y_X^{\rm final} \approx Y_{X,0}/S_{\rm ratio}$, where $S_{\rm ratio} = S_{\rm final}/S_0$ is the ratio of the final to initial comoving entropy, $S = s\,a^3$, computed numerically along the integration of Eqs.~\eqref{eq:dlnTdN}--\eqref{eq:dlnrhoXdN}, terminated once $\rho_Y/\rho_{\rm tot}$ drops below a small threshold, ensuring that the mediator has fully decayed and the entire entropy injection has been captured.

\subsection{Approaching the Thermalization floor}\label{sm:thermfloor}

\noindent  A key design choice of
the numerical implementation outlined so far is to treat the mediator as stable during the Boltzmann evolution, with the portal coupling, $\varepsilon$, entering only in the late-time background stage. This separation allows us solve the stiff Boltzmann system once for a given set of hidden-sector parameters, and then scan rapidly over $\varepsilon$ through the much cheaper background evolution to find the value that  produces the observed relic abundance. For sufficiently large $\varepsilon$, however, $\Gamma_Y$ becomes comparable to $H$ during freeze-out and this separation no longer holds. 

To extend our results into this regime, we employ a joint solver that evolves the hidden sector together with the portal-induced $Y \leftrightarrow \mathrm{SM}$ exchanges throughout freeze-out. Concretely, this affects two of the seven coefficient functions of the Boltzmann system in Eqs.~\eqref{eq:dmuX}-\eqref{eq:dlnxi}, namely those governing the mediator number density and the hidden-sector temperature,
\begin{align}
    C &\to C - \frac{\tilde g}{H_{\rm tot}\,x}\,\Gamma_Y\,\Big(1 - \frac{n_Y^{\rm eq,SM}}{n_Y}\Big)\,, \label{eq:C_decay}\\
    E &\to E - \frac{\tilde g}{H_{\rm tot}\,x}\,\frac{\Gamma_Y\,(B_1^Y\,n_Y - B_1^{Y,\,\mathrm{SM}}\,n_Y^{\rm eq,SM})}{\tilde{\mathcal{D}}}\,, \label{eq:E_decay}
\end{align}
where $n_Y^{\rm eq,SM}$ is the $Y$ equilibrium number density and $B_1^{Y,\,\mathrm{SM}}$ is energy-per-particle coefficient evaluated at the SM temperature. The first term combines the forward decay, $Y \to \mathrm{SM}$, that depletes the mediator yield with the inverse decay, $\mathrm{SM} \to Y$, that repopulates it. The second is the corresponding net energy flow, removing the average energy per particle from the hidden sector for each forward decay and replenishing it for each inverse decay. When both rates exceed the Hubble rate, $n_Y \to n_Y^{\rm eq,SM}$ and the hidden sector thermalizes with the SM bath, recovering the standard WIMP picture.

For slightly lower values of the portal coupling, $\varepsilon$, only the forward decay, $Y \to \mathrm{SM}$, remains efficient, due to the cannibal-heating of the hidden sector relative to the SM which leads to a suppression of the inverse-decay channel by $n_Y^{\rm eq,\, SM}/n_Y\ll1$. This in turns effects the freeze-out process by limiting the $YY \to X\bar X$ backreaction due to the persistent $Y$ decays. 

Finally, for even lower values of $\varepsilon$, namely $\varepsilon \lesssim 10^{-10}$, both rates remain below the Hubble rate throughout freeze-out and the secluded picture of Sec.~\ref{sm:solver} --- which is the regime of primary interest in this work --- is recovered. We have explicitly checked that the two evolutions agree at the sub-percent level across our parameter space.

\section{Hypercharge Portal: Cross Sections and Rates}
\label{sm:hyper}
 
\noindent  This section collects the hidden-sector rates, mediator decay width, and spin-independent DM--nucleon cross section of the hypercharge portal model analyzed in the main text.
 
\subsection{Hidden-Sector Cross Sections}\label{sm:hyper_dark}
 
\noindent  In this model, the thermally averaged DM annihilation cross section takes the form:
\begin{equation}
   \frac{1}{2} \langle\sigma v\rangle_{X\bar X \to Z^\prime Z^\prime } =a + \frac{6 b}{x_X},
    \label{eq:VP_sv_full}
\end{equation}
where $x_X \equiv m_X/T_h$ and the overall factor of $1/2$ accounts for the Dirac nature of $X$, avoiding the double counting of $X \bar X$ pairs in the initial state when the Boltzmann equation is written in terms of the total number density, $n_{X+\bar X}\equiv n_X + n_{\bar X}$. The coefficients $a$ and $b$ are respectively the s-wave and p-wave contributions, given by: 
\begin{align}
    a &= \frac{2\pi\,\alpha_X^2}{m_X^2}\,\frac{(1-r^2)^{3/2}}{(2-r^2)^2}\,, \label{eq:VP_s}\\
    b &= \frac{\pi\,\alpha_X^2}{12\,m_X^2}\,\frac{\sqrt{1-r^2}\,(24 + 28 r^2 - 36 r^4 + 17 r^6)}{(2-r^2)^4}\,, \label{eq:VP_p}
\end{align}
where $r \equiv m_Y/m_X$.
The s-wave coefficient, $a$, coincides with the expression quoted in the main text, i.e. Eq.~\eqref{eq:sv_swave}, while the p-wave contribution is important at high temperatures but becomes Boltzmann-suppressed at and after freeze-out, for $x_X\gg 1$.

The mediator, $Z^\prime$, is kept in chemical equilibrium throughout freeze-out by three $3 \to 2$ number-changing (cannibal) reactions within the hidden sector. Following Ref.~\cite{Berlin:2016gtr}, we parametrize their thermally averaged cross sections through dimensionless coefficients, $\Delta_{1,2,3}$, so that the corresponding rates per mediator read
\begin{align}
    \Gamma_{Z^\prime Z^\prime Z^\prime  \to Z^\prime Z^\prime } &=n_{Z^\prime }^2\,\langle\sigma v^2\rangle_{Z^\prime }\, \nonumber \\
    &= \,n_{Z^\prime }^2\,\Delta_1\,\frac{\alpha_X^5\,T_h^7}{m_X^{12}}\,,\label{eq:VP_sv2_YYY} \\
    \Gamma_{Z^\prime \bar XX \to \bar X X} &=\frac{1}{2}n_{X}^2\, \langle\sigma v^2\rangle_{XX} \nonumber \\
    &= \frac{1}{2}n_X^2\,\Delta_2\,\frac{\alpha_X^3}{m_X^5}\,, \label{eq:VP_sv2_YXX}\\
    \Gamma_{Z^\prime Z^\prime X \to Z^\prime X} &=n_{Z^\prime }\,n_X \, \langle\sigma v^2\rangle_{Z^\prime X}\nonumber \\
    &=n_{Z^\prime }\,n_X\,\Delta_3\,\frac{\alpha_X^3}{m_X^5}\,.\label{eq:VP_sv2_YYX}
\end{align}
Of the three, the most suppressed is the purely-mediator channel, $Z^\prime Z^\prime Z^\prime \to Z^\prime Z^\prime $, which proceeds through a one-loop diagram and is therefore further suppressed by $\alpha_X^2$ relative to the other two, with an additional explicit temperature dependence that reduces its contribution at late times. It is numerically negligible throughout the parameter space considered, and we retain it only for completeness, fixing its coefficient to the phenomenological value $\Delta_1 = 0.5$ adopted in Ref.~\cite{Berlin:2016gtr}. The remaining two channels, $Z^\prime Z^\prime X\to Z^\prime X$ and $Z^\prime XX\to XX$, are both tree-level and share the same $\sim \alpha_X^3/m_X^5$ scaling so that, up to the $\sim \mathcal{O}(1-10)$ ratio $\Delta_3/\Delta_2$, their relative importance is set by the hidden-sector number densities through $n_{Z^\prime }/n_X$. Since $m_{Z^\prime }<m_X$, the mediator is less Boltzmann-suppressed than the DM at late times, so that $Z^\prime Z^\prime X\to Z^\prime X$ is generically the dominant cannibal channel through freeze-out. 

We compute the coefficients $\Delta_2$ and $\Delta_3$ using \texttt{FeynCalc}~\cite{Mertig:1990an, Shtabovenko:2016sxi, Shtabovenko:2020gxv, Shtabovenko:2023idz}, specifically in the non-relativistic limit, which is the regime of interest for our purposes, as these rates only begin to affect the dynamics once $Z^\prime $ starts to depart from chemical equilibrium. We find:
\begin{align}
    \Delta_2 &=\frac{\pi^{2}\,(4+r)^{3/2}\,\mathcal{N}_2(r)}{6\,r^{7/2}\,(2+r)^{3}\,(2+r-r^2)^{2}}\,, \label{eq:VP_Delta2} \\
    \Delta_3 &= \frac{4\pi^{2}\,\sqrt{(2+r)(2+3r)}\,\mathcal{N}_3(r)}{\sqrt{3}\,r\,(1+2r)^{3}\,(2+r)^{4}\,\bigl(r^{3}-r^{2}-4r-2\bigr)^{2}}\,, \label{eq:VP_Delta3}
\end{align}
where we have defined the quantities
\begin{align}
    \mathcal{N}_2(r) &=
    31\,r^{6} + 72\,r^{5} + 38\,r^{4} - 96\,r^{3} \nonumber \\
    &+ 536\,r^{2} - 384\,r + 256\,,\label{eq:VP_N2} \\
    \mathcal{N}_3(r) &=  195\,r^{8} + 1156\,r^{7} + 4670\,r^{6} \nonumber \\
    &+ 9444\,r^{5} + 12214\,r^{4} + 11192\,r^{3} \nonumber \\
    &+ 6732\,r^{2} + 2272\,r + 320
    \label{eq:VP_N3}
\end{align}

\subsection{Mediator Decay Width}
\label{sm:hyper_decay}

\noindent  The mediator, $Z^\prime$, couples to SM fermions through its kinetic mixing with the hypercharge gauge boson, which induces $\varepsilon$-suppressed couplings to any SM fermion, $f$, with electric charge, $Q_f$, or weak hypercharge, $Y_L$, $Y_R$.  The relevant vector and axial couplings are $g_{fV} = (g_R + g_L)/2$ and $g_{fA} = (g_R - g_L)/2$, with~\cite{Hoenig:2014dsa, Berlin:2016vnh}:
\be
    g_{L,R} = \varepsilon\,\frac{m_{Z^\prime }^2\,g_Y^{\rm SM}\,Y_{L,R} - m_Z^2\,g_W\,\sin\theta_W\,\cos\theta_W\,Q_f}{m_Z^2 - m_{Z^\prime }^2},~~~~
\ee
where $g_Y^{\rm SM}$ and $g_W$ are the hypercharge and weak gauge couplings, respectively, $\theta_W$ is the Weinberg angle, and $m_Z$ the $Z$-boson mass.

These couplings determine the total $Z^\prime $ decay width, obtained by summing the partial widths of all kinematically accessible fermion-pair channels~\cite{ Hooper:2019xss, Hu:2025thq},
\be
    \Gamma_{Z^\prime } &=& \sum_{f:\,m_{Z^\prime } > 2 m_f} \frac{N_c^f\,m_{Z^\prime }}{12\pi}\,\sqrt{1-\frac{4 m^2_f}{m^2_{Z^\prime }}}\, \nonumber \\
    &&\times \Bigg[g_{fV}^2\,\left(1+\frac{2 m^2_f}{m^2_{Z^\prime }}\right)+ g_{fA}^2\,\left(1-\frac{4 m^2_f}{m^2_{Z^\prime }}\right)\Bigg]\,,~~~~~~~~
    \label{eq:VP_Gamma_tot}
\ee
where $N_c^f$ is the color multiplicity of the SM fermion ($3$ for quarks, $1$ for leptons) with mass $m_f$.

\subsection{Spin-Independent DM--Nucleon Cross Section}\label{sm:hyper_si} 

\noindent  The same kinetic mixing that controls the $Z^\prime$ decay also generates a tree-level DM--nucleon elastic scattering amplitude, mediated by the coherent exchange of the two mass eigenstates, $Z$ and $Z^\prime$, after electroweak symmetry breaking~\cite{Alenezi:2025kwl, Hu:2025thq}. In the parameter space of interest here, i.e. for $m_{Z^\prime}/m_X \gtrsim 0.1$, recoil-energy effects on the scattering are negligible~\cite{Alenezi:2025kwl}, and the corresponding spin-independent DM--nucleon cross section reads simply as~\cite{Cline:2014dwa, Alenezi:2025kwl, Hu:2025thq}
\begin{align}
    \sigma_{\rm SI} = \frac{\mu^2}{\pi}\Bigg[&\sum_{i\in\{Z,Z'\}}\frac{V_X^i\,\mathcal{Z}\,(2 V_u^i + V_d^i)}{\mathcal{A}\,m_i^2} \nonumber \\
    &+\frac{V_X^i\, (\mathcal{A}-\mathcal{Z})\,(V_u^i + 2 V_d^i)}{\mathcal{A}\,m_i^2}\Bigg]^{2},
    \label{eq:VP_sigma_SI}
\end{align}
where $\mathcal{Z}$ and $\mathcal{A}$ are the atomic and mass numbers of the target nucleus, $m_N = (m_p + m_n)/2\approx 0.939\,{\rm GeV}$ is the average nucleon mass, and $\mu \equiv m_X m_N/(m_X + m_N)$ is the DM--nucleon reduced mass. The DM couplings, $V_X^Z$ and $V_X^{Z^\prime}$,
 are set by the $Z-Z^\prime$ mixing angle, $\eta$, defined as~\cite{Hu:2025thq}
\begin{equation}
    \tan(2\eta) = \frac{2\,\varepsilon\,\sin\theta_W}{1 - m_{Z'}^2/m_Z^2}\,,
    \label{eq:VP_eta}
\end{equation}
and are given by $V_X^Z=g_X\sin\eta$ and $V_X^{Z^\prime}=g_X\cos\eta$. On the other hand, the corresponding couplings to the SM quarks, $q=\{u,\,d\}$, receive both an electromagnetic and a neutral-current contribution,
\begin{align}
    V_q^Z =& \sqrt{4\pi\alpha_{\rm EM}}\,\varepsilon\,s_\eta\,Q_q \nonumber \\
    &+ \frac{\sqrt{4\pi\alpha_{\rm EM}}}{s_W\,c_W}\,\big(c_\eta + \varepsilon\,s_W\,s_\eta\big)\,\big(T_3^q - s^2_W\,Q_q\big)\,, \\
    V_q^{Z'} &= \sqrt{4\pi\alpha_{\rm EM}}\,\varepsilon\,c_\eta\,Q_q \nonumber \\
    &+ \frac{\sqrt{4\pi\alpha_{\rm EM}}}{s_W\,c_W}\,\big(s_\eta - \varepsilon\,s_W\,s_\eta\big)\,\big(T_3^q - s^2_W\,Q_q\big)\,,
\end{align}
where $\left( s_W,\, c_W,\,s_\eta,\,c_\eta\right)=\left(\sin\theta_W,\,\cos\theta_W,\,\sin\eta,\,\cos\eta\right)$, $Q_q$
and $T_3^q$ are the electric charge and weak isospin of quark, $q$,  and $\alpha_{\rm EM}\simeq 1/137$ is the fine-structure constant.

We compare the resulting cross section against the current limits from LUX-ZEPLIN (LZ)~\cite{LZ:2024zvo} and XENONnT~\cite{XENON:2025vwd}, both of which employ xenon targets ($\mathcal{Z} = 54$, $\mathcal{A} = 131$).  Two features of Eq.~\eqref{eq:VP_sigma_SI} are worth highlighting. First, the mixing angle defined in Eq.~\eqref{eq:VP_eta} diverges at $m_{Z'} = m_Z$, causing the vector couplings, $V_q^{Z,Z'}$, to blow up on resonance. This pole would be regularized by the finite $Z'$ width, which we are neglecting here, but the resonance enhancement remains sizable in the immediate vicinity of the $Z$ mass. Second, the coherent sum of $Z$ and $Z'$ exchange in Eq.~\eqref{eq:VP_sigma_SI} admits an exact cancellation below the pole, where the two amplitudes interfere destructively. For a xenon target, this blind spot occurs at
\begin{equation}
    m_{Z'}^{\cancel{\sigma}} \simeq 64.3~\mathrm{GeV} \approx 0.705\,m_Z\,,
    \label{eq:VP_blind_spot}
\end{equation}
at which $\sigma_{\rm SI}$ vanishes identically, to leading order in $\varepsilon$. The location of this zero depends on the target nucleus through the $(\mathcal{Z},\mathcal{A})$-weighted quark couplings in Eq.~\eqref{eq:VP_sigma_SI}, but not on any hidden-sector parameter.

\section{Constraints from Big Bang Nucleosynthesis}\label{sm:bbn}

\noindent  Hidden-sector freeze-out leaves behind a sizable comoving abundance of mediators, $Z^\prime$, which, for sufficiently small values of $\varepsilon$, can survive sufficiently long to decay during or after BBN. These decays inject hadronic energy into the primordial plasma and can thereby disrupt the successful predictions of the light-element abundances. The resulting constraints take the form of an upper bound on the mediator energy yield per decay, $\zeta \equiv m_{Z'}\,Y_{Z'}$. 

Here, we adopt the bounds derived in Ref.~\cite{Kawasaki:2017bqm}, tabulated for two representative hadronic injection channels ($u\bar u$ and $b\bar b$) over six reference masses spanning $30\,\mathrm{GeV}$ to $1000\,\mathrm{TeV}$. The bound is applied to each kinematically accessible decay channel independently, weighted by its branching ratio. Leptonic channels carry no hadronic constraint in the short-lifetime regime relevant here, while the remaining hadronic final states (up- and down-type quarks, gluons, and electroweak bosons) produce similar pion-dominated showers and are grouped under the $u\bar u$ curves of Ref.~\cite{Kawasaki:2017bqm}, so that only the relative weight of $b\bar b$ against this $u\bar u$-like combination matters.  A two-dimensional log--log interpolation is performed on this grid in the $(\log_{10} m_{Z'},\,\log_{10}\tau_{Z'})$ plane. Outside the tabulated mass range, the bound is extended via the following power-law extrapolation:
\begin{equation}
    \zeta_{\max}(m_{Z'}) = \zeta_{\max}(m_{\rm anchor})\,(m_{Z'}/m_{\rm anchor})^\alpha\,,
    \label{eq:BBN_extrap}
\end{equation}
where the slope, $\alpha$, is extracted as the local power-law index between adjacent tabulated masses at fixed $\tau_{Z'}$, and $m_{\rm anchor}$ is the closest tabulated mass to the extrapolation region (i.e. $1000\,\mathrm{TeV}$ above the grid and $30\,\mathrm{GeV}$ below it). Above $1000\,\mathrm{TeV}$ both channels converge to $\alpha_{\rm high} \simeq 0.70$, a value that follows from the QCD fragmentation multiplicity, $N_{\rm had} \sim m_{Z'}^{0.3}$~\cite{Kawasaki:2017bqm}. This implies that heavier particles produce more hadrons per decay, so fewer decays are needed to inject the same hadronic energy, and thus the bound on the total energy yield relaxes as $\zeta_{\max} \propto m_{Z'}^{1-0.3}$. Below $30\,\mathrm{GeV}$ the slopes flatten to $\alpha_{\rm low} = 0.50$ for $u\bar u$ and $0.34$ for $b\bar b$, reflecting the entry into a regime in which the energy per nucleon in the shower drops below the threshold for efficient hadrodissociation, and the bound tightens more slowly with mass than the multiplicity argument alone would predict. The particularly low $b\bar b$ slope reflects an additional suppression due to the proximity to the $b$-quark threshold, which sharply reduces shower multiplicity relative to $u\bar u$. 

In practice, however, for the large mediator abundances characteristic of the hidden-sector scenarios considered in this work, the constraint reduces to a single lifetime threshold, $\tau_{Z'} \lesssim 0.1$--$0.2\,\mathrm{s}$, set by the requirement that charged pions from the hadronic shower thermalize in the electromagnetic plasma before they can drive $p \leftrightarrow n$ interconversion~\cite{Kawasaki:2017bqm}.

As a corollary, the final result is essentially insensitive to the detailed shape of the $\zeta_{\max}(\tau_{Z'})$ curve, and hence to the interpolation, extrapolation, and channel-grouping choices entering the implementation above.

\section{Results for Other Portals}\label{sm:other}

\noindent  The hidden-sector framework analyzed in the main text extends naturally to a broad class of DM models communicating with the SM through a portal interaction. Here we present the results obtained for three additional benchmark portals: the Higgs portal, the $B-L$ portal, and the baryon portal. Many other choices are possible, but these three suffice to illustrate the robustness of our findings across qualitatively different mediator structures and SM coupling patterns.

\subsection{The Higgs Portal}\label{sm:higgs}

\noindent  In the Higgs portal scenario, a real scalar ($g_Y=1$), $\phi$,  couples to the SM Higgs at tree level through the mass mixing parametrized by a small mixing angle, $\varepsilon$. For the DM candidate, $X$, we consider a Majorana fermion ($g_X = 2$) that is coupled to the scalar mediator, $\phi$, through hidden-sector Yukawa interactions, following Ref.~\cite{Berlin:2016gtr}:
\begin{equation}
    \mathcal{L} \supset -\frac{\lambda_s}{2}\,\phi\,\bar{X}X - \frac{i\lambda_p}{2}\,\phi\,\bar{X}\gamma_5 X\,,
    \label{eq:HP_Lyuk}
\end{equation}
where $\lambda_s$ and $\lambda_p$ are the scalar and pseudoscalar Yukawa couplings, respectively.

\subsubsection{Cross Sections and Rates}

\noindent  In this model, the thermally averaged DM annihilation cross section takes the form
\begin{equation}
    \langle\sigma v\rangle_{X\bar X \to \phi\phi} = a + \frac{6\,b}{x_X}\,,
\end{equation}
where the coefficients $a$ and $b$ represent the
s-wave and p-wave contributions, respectively, and are given by
\begin{align}
    a =& \frac{2\sqrt{1-r^2}\,\lambda_s^2\,\lambda_p^2}{\pi\,m_X^2\,(r^2 - 2)^2},
    \label{eq:HP_swave} \\
    b =&  \frac{1}{12\,m_X^2\,\pi\,\sqrt{1-r^2}\,(r^2-2)^4}\Big[
    -2(r^2-1)^3\,\lambda_p^4
    \nonumber \\
    &+ 3(r^6 - 8r^4 + 20r^2 - 12)\,\lambda_s^2\,\lambda_p^2 \nonumber \\
    &+ 2(-2r^6 + 10r^4 - 17r^2 + 9)\,\lambda_s^4\Big].\label{eq:HP_pwave}
\end{align}
 Note that the s-wave contribution vanishes whenever either Yukawa coupling is taken to zero, so both $\lambda_s$ and $\lambda_p$ must be nonzero in order to obtain a non-suppressed annihilation rate.

The mediator, $\phi$, is kept in chemical equilibrium throughout freeze-out by the same three $3 \to 2$ cannibal reactions as in the hypercharge portal, with rates again parametrized through dimensionless coefficients, $\Delta_{1,2,3}$, that now depend on both $r$ and the coupling ratio, $\rho \equiv \lambda_p/\lambda_s$,
\begin{align}
    \Gamma_{\phi\phi X \to \phi X} &= \Delta_3(r,\rho)\,\frac{\lambda_s^3\,\lambda_p^3}{m_X^5}\,n_\phi\,n_X\,, \\
    \Gamma_{\phi X X \to X X}      &= \Delta_2(r,\rho)\,\frac{\lambda_s^3\,\lambda_p^3}{m_X^5}\,n_X^2\,, \\
    \Gamma_{\phi\phi\phi \to \phi\phi} &= \Delta_1(r,\rho)\,\frac{\lambda_s^5\,\lambda_p^5}{m_X^5}\,n_\phi^2\,.
\end{align}

The hierarchy among the three channels mirrors that of the hypercharge case. Specifically, $\phi\phi X \to \phi X$ dominates throughout freeze-out, while $\phi\phi\phi \to \phi\phi$ proceeds through a one-loop diagram (a closed $X$ loop with five external $\phi$ legs) and is generically suppressed relative to the other two. We compute $\Delta_2$ and $\Delta_3$ exactly at tree level using \texttt{FeynCalc}. $\Delta_1$ is instead computed in the heavy-fermion limit, $r \gg 1$, which is the only regime in which the loop suppression of this channel can be compensated, as the tree-level processes are then Boltzmann-suppressed by the depleted $X$ abundance while $\Delta_1$ requires $X$ only as a virtual particle inside the loop~\cite{Berlin:2016gtr}. 
The resulting expressions read as follows:
\begin{align}
      \Delta_1 &= \frac{\sqrt{5}}{3\pi^5}\,\frac{(3 + 10\rho^2 + 15\rho^4)^2}{r^3\,\rho^3}\,,
    \label{eq:HP_Delta1} \\
    \Delta_2 &= \frac{\sqrt{r+4}}{64\pi\,r^{5/2}\,(r+2)^3\,(2+r-r^2)^2} \nonumber \\
    &\times \Big[
    r^5(r+8)(1+\rho^2)^3
    + 4r^4(2 + 21\rho^2 + 20\rho^4 + \rho^6) \nonumber \\
    &- 16r^3(4 - 7\rho^2 - 6\rho^4 + 5\rho^6) - 16r^2(7 - 6\rho^2 \nonumber \\
    &- 25\rho^4 + 4\rho^6) 
    + 128r(1 + 9\rho^2 + 14\rho^4 + 2\rho^6) \nonumber \\
    &+ 256(1 + 2\rho^2)^2
    \Big]\,,
    \label{eq:HP_Delta2} \\
 \Delta_3&= \frac{9\sqrt{3}\,\sqrt{3r^2+8r+4}}{64\pi\,r\,(1+2r)^3\,(2+r)^4\,(2+4r+r^2-r^3)^2} \nonumber \\
    &\times \Big[
    r^2(r+2)^5(3r+2) + r(9r^5 + 16r^4 + 180r^3 + 400r^2 \nonumber \\
    &+ 288r + 64)(r+2)^2\rho^2
    + 3r^4(2+2r-r^2)^2\rho^6 \nonumber \\
    &+ (9r^8 + 8r^7 + 140r^6 + 888r^5 + 2220r^4
    + 3360r^3 \nonumber \\
    &+ 3024r^2 + 1408r+ 256)\rho^4
    \Big]\,.
    \label{eq:HP_Delta3} 
\end{align}

Having specified the hidden-sector dynamics, we now turn to the connection to the SM. The mediator, $\phi$, decays into SM particles through its mixing with the SM Higgs, with all partial widths proportional to $\sin^2\varepsilon \approx \varepsilon^2$ in the small-mixing limit. We therefore write the total width as $\Gamma_\phi = \varepsilon^2\,\Gamma_{\rm SM}(m_\phi)$, where $\Gamma_{\rm SM}(m_\phi)$ is the SM Higgs decay width evaluated at $m_h = m_\phi$. 

Because the Higgs phenomenology spans many decades in mass with distinct kinematic regimes, $\Gamma_{\rm SM}$ is computed through a hybrid procedure across three mass windows. Below $5\,\mathrm{GeV}$, where leptonic and low-multiplicity hadronic final states dominate, we use the \texttt{scalar\_portal} package~\cite{Winkler:2018qyg} which combines a dispersive analysis below $2\,\mathrm{GeV}$ with perturbative QCD using running quark masses up to $5\,\mathrm{GeV}$. Between $5$ and $500\,\mathrm{GeV}$, we use the \texttt{HDECAY} package~\cite{Djouadi:1997yw,Djouadi:2018xqq}, which includes higher-order QCD corrections, off-shell $WW^\ast/ZZ^\ast$ decays, and loop-induced $gg$ and $\gamma\gamma$ channels. Above $500\,\mathrm{GeV}$, we revert to tree-level kinematics as unresummed Sudakov logarithms make \texttt{HDECAY} unreliable. For $m_\phi > 2 m_h$, an additional decay channel arises in the singlet extension~\cite{Berlin:2016gtr}:
\begin{equation}
    \Gamma(\phi\to hh) =\varepsilon^2 \frac{(2m_\phi^2 + m_h^2)^2}{128\pi\,m_\phi\,v_H^2}\,\sqrt{1 - \frac{4m_h^2}{m_\phi^2}},
\end{equation}
where $v_H = 246.22\,{\rm GeV}$ is the Higgs vacuum expectation value.

Finally, the spin-independent DM--nucleon scattering amplitude relevant for direct detection proceeds via the $t$-channel exchange of the two mixed scalars, $h$ and $\phi$. The resulting cross section is
\begin{equation}
    \sigma_{\rm SI} =\frac{4\mu^2\lambda^2_s\,\varepsilon^2\,f^2_N\,m^2_N}{\pi \,v^2_H}\left(\frac{1}{m_\phi^2} - \frac{1}{m_h^2}\right)^{2}\,,
    \label{eq:HP_sigma_SI}
\end{equation}
where $f_N \simeq 0.30$ is the Higgs--nucleon form factor, $\mu$ is the DM--nucleon reduced mass, and the factor of $4$ reflects the Majorana nature of $X$. The result is isospin-universal and independent of the target nucleus, since the coupling to nucleons proceeds entirely through the Higgs--nucleon vertex. The propagator difference vanishes at $m_\phi = m_h$, where the two amplitudes interfere destructively and $\sigma_{\rm SI} = 0$ identically. At this blind spot the model is invisible to direct detection regardless of the hidden-sector parameters.

\subsubsection{Results}

 \noindent  In the top panel of Fig.~\ref{fig:s1}, we show the analogue of Fig.~\ref{fig:3} in the main text for the case of the Higgs portal, evaluated for $\lambda_s = \lambda_p = 0.05$ and for $\lambda_s = \lambda_p = 0.1$. The qualitative picture parallels that of the hypercharge portal, with direct detection bounding the relic line from above and BBN from below, leaving a large portion of viable parameter space at small $\varepsilon$ where the model is currently unconstrained. The low-velocity annihilation cross section transitions from WIMP-like values at $\varepsilon \gtrsim 10^{-9}$ to progressively smaller values at lower $\varepsilon$, as entropy injection from late time $\phi$ decays dilutes the DM relic abundance.
 
\subsection{The $B-L$ Portal}
\label{sm:bml}

\noindent  In the $B-L$ portal scenario, the hidden-sector vector, $Z'$, mixes kinematically with the $U(1)_{B-L}$ gauge boson, $Z_{B-L}$. As the DM particle, $X$, we consider a Dirac fermion that couples to the $Z'$ in the same way as in the hypercharge portal. The hidden-sector dynamics (namely DM annihilation, $X\bar X \to Z'Z'$, and the three $3 \to 2$ $Z^\prime$ cannibal channels) are thus identical to those of the hypercharge portal, as described in Sec.~\ref{sm:hyper_dark}.

\subsubsection{Connection to the Standard Model}

\noindent  In contrast to the hypercharge portal, here the kinetic mixing induces purely vectorial couplings of the $Z'$ to SM fermions, $f$, namely~\cite{Hooper:2019xss}:
\begin{align}
    g_{fV} = \varepsilon\,g_{B-L}\,(B-L)_f\,\left|\frac{m_{Z_{B-L}}^2 + m_{Z'}^2}{m_{Z_{B-L}}^2 - m_{Z'}^2}\right|\,, 
    \label{eq:BL_coupling}
\end{align}
where $g_{B-L}$ is the $U(1)_{B-L}$ gauge coupling, $m_{Z_{B-L}}$ the corresponding gauge boson mass, and $(B-L)_f = +1/3$ for quarks and $-1$ for leptons. Throughout this work we adopt $m_{Z_{B-L}} = 10\,\mathrm{TeV}$, which is sufficiently above the $Z'$ mass range of interest that the kinematic factor in Eq.~\eqref{eq:BL_coupling} is effectively unity and the coupling reduces to $g_{fV} \approx \varepsilon\,g_{B-L}\,(B-L)_f$. 

The total $Z'$ decay width retains the same functional form as in the hypercharge portal (see Eq.~\eqref{eq:VP_Gamma_tot}), evaluated with $g_{fA} = 0$ and summed over all kinematically accessible SM fermions, including leptons.

For direct detection, the universal $(B-L)_q = 1/3$ quark charge makes the resulting spin-independent DM--nucleon cross section isospin-universal, with protons and neutrons receiving equal contributions, giving
\begin{equation}
    \sigma_{\rm SI} \approx\frac{4 \mu^2\alpha_X \varepsilon^2\, g_{B-L}^2}{m^4_{Z^\prime}}\,,
    \label{eq:BL_sigma_SI}
\end{equation}
where $\mu$ is the DM--nucleon reduced mass. Unlike in the hypercharge and Higgs portal models, only a single mediator contributes in this case, so no blind spot arises.

\subsubsection{Results}
\label{sec:BL_results}

\noindent  In the central panel of Fig.~\ref{fig:s1}, we show the analogue of Fig.~\ref{fig:3} in the main text for the case of the $B-L$ portal, evaluated at $\alpha_X = 10^{-2}$ and $10^{-3}$. The structure of the bounds mirrors closely that of the hypercharge portal, with the relic line again bounded from above by direct detection and from below by BBN, with a sizeable region at small portal coupling remaining free of any current bound. The predicted low-velocity annihilation cross section tracks the WIMP value at $\left(\varepsilon \,g_{B-L}\right)\gtrsim 10^{-11}$ and falls below it at lower $\varepsilon$, as the dilution from late time $Z'$ decays takes over.

\subsection{The Baryon Portal}\label{sm:baryon}
\label{sec:SM_B}

\noindent  In the Baryon portal scenario, the hidden-sector vector $Z'$ mixes kinetically with the $U(1)_B$ gauge boson, $Z_B$~(see e.g.~\cite{Carone:1994aa, FileviezPerez:2010gw, Duerr:2013dza}). This is not to be confused with baryon portal models involving operators of the form, $u dd$. Again, the DM candidate, $X$, is a Dirac fermion coupled to the $Z'$ in the same way as in the hypercharge portal model, so the hidden-sector dynamics (DM annihilation, $X\bar X \to Z'Z'$, and the three $3 \to 2$ $Z'$ cannibal channels) are again identical to those of the hypercharge portal, as described in Sec.~\ref{sm:hyper_dark}.

\subsubsection{Connection to the Standard Model}

\noindent  Kinetic mixing again induces purely vectorial couplings of the $Z'$ with SM fermions, $f$, but now limited to quarks alone~\cite{Hooper:2019xss},
\begin{align}
    g_{fV} = \varepsilon\,g_B\,B_f\,\left|\frac{m_{Z_B}^2 + m_{Z'}^2}{m_{Z_B}^2 - m_{Z'}^2}\right|\,,
    \label{eq:B_coupling}
\end{align}
where $g_B$ is the $U(1)_B$ gauge coupling, $m_{Z_B}$ the corresponding gauge boson mass, and $B_f = +1/3$ for quarks and $0$ for leptons. As in the $B-L$ case, we adopt $m_{Z_B} = 10\,\mathrm{TeV}$, which is sufficiently above the $Z'$ mass range of interest that the kinematic factor in Eq.~\eqref{eq:B_coupling} is effectively unity and the coupling reduces to $g_{fV} \approx \varepsilon\,g_B\,B_f$. The total $Z'$ decay width then retains the same functional form in Eq.~\eqref{eq:VP_Gamma_tot}, but the sum runs only over kinematically accessible quark--antiquark final states, since leptons are uncharged under $U(1)_B$.

For direct detection, the universal $B_q = 1/3$ quark charge again makes the resulting spin-independent DM--nucleon cross section isospin-universal, taking the same form as in Eq.~\eqref{eq:BL_sigma_SI} for the $B-L$ portal under the replacement $g_{B-L} \to g_B$:
\begin{equation}
    \sigma_{\rm SI} \approx \frac{4\,\mu^2\,\alpha_X\,\varepsilon^2\,g_B^2}{m_{Z'}^4}\,.
    \label{eq:B_sigma_SI}
\end{equation}
Finally, as for the $B-L$ portal, only a single mediator contributes, so the destructive interference responsible for the hypercharge and Higgs portal 
blind spots are absent in this case.

\subsubsection{Results}
\label{sec:B_results}

\begin{figure*}[ht!]
    \centering
    \begin{minipage}{0.9\linewidth}
        \centering
        \includegraphics[width=.95\linewidth]{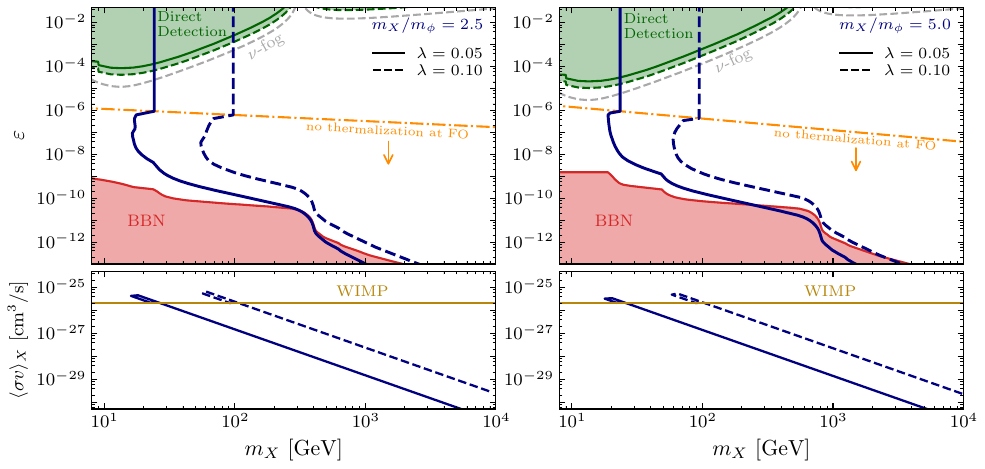}\\
        (a) Higgs Portal
    \end{minipage}
    
    \vspace{0.5em}
    
    \begin{minipage}{0.9\linewidth}
        \centering
        \includegraphics[width=.95\linewidth]{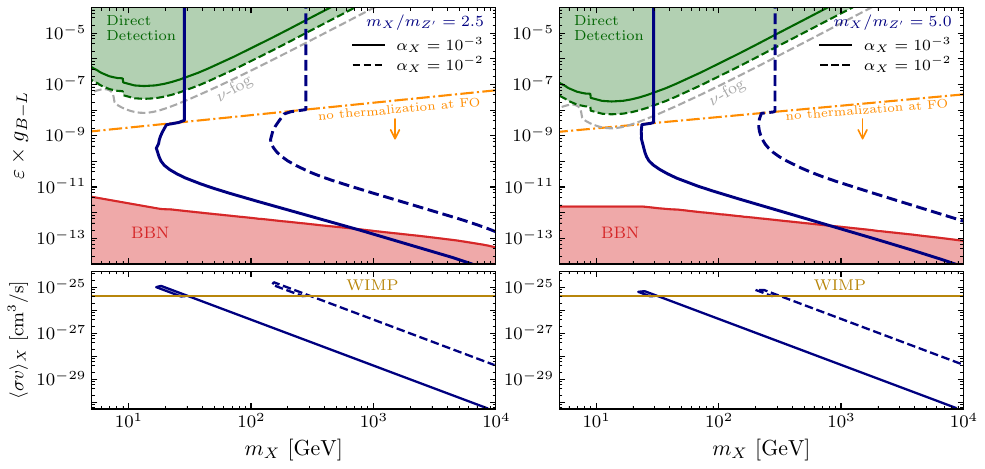}\\
        (b) $B-L$ Portal
       
    \end{minipage}
    
    \vspace{0.5em}
    
    \begin{minipage}{0.9\linewidth}
        \centering
        \includegraphics[width=.95\linewidth]{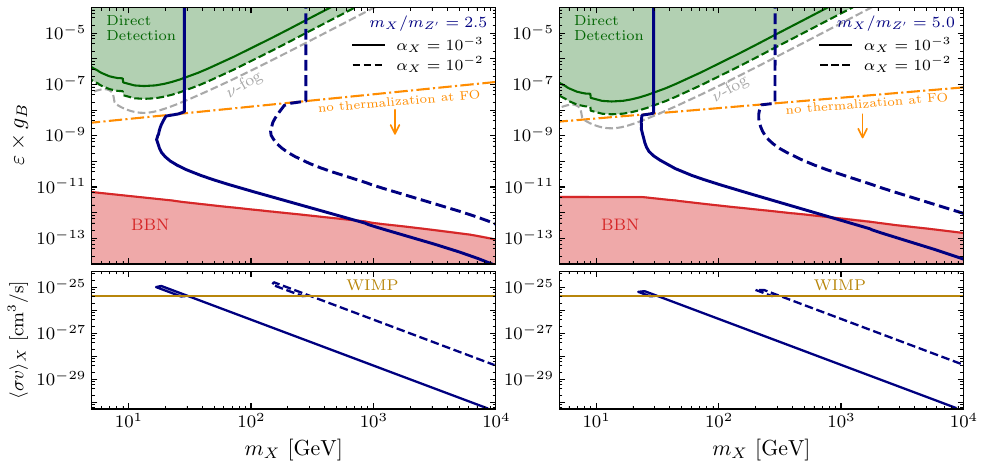}\\
        (c) Baryon Portal
    \end{minipage}
    
    \caption{As in Fig.~\ref{fig:3}, but for each of the three additional portal scenarios considered.}
    \label{fig:s1}
\end{figure*}

\noindent  In the lower panel of Fig.~\ref{fig:s1}, we show the analogue of Fig.~\ref{fig:3} in the main text for the case of the Baryon portal, again at $\alpha_X = 10^{-2}$ and $10^{-3}$. The phenomenology is essentially identical to that of the $B-L$ portal, with the same overall structure of bounds and the same minor suppression of the predicted annihilation cross section at $\left(\varepsilon\, g_B\right)\lesssim 10^{-11}$.

\end{document}